\documentclass[12pt]{article}

\usepackage{graphics}
\usepackage{amssymb}

\renewcommand{\theequation}{\arabic{section}.\arabic{equation}}

\newcommand{\D}{\mathrm{D}}

\newcommand{\R}{\mathbb{R}}
\newtheorem{claim}{Claim}[section]
\newtheorem{theorem}[claim]{Theorem}
\newtheorem{lemma}[claim]{Lemma}
\newtheorem{remark}[claim]{Remark}
\newtheorem{remarks}[claim]{Remarks}
\newenvironment{proof}[1][Proof]{\textsl{#1.} }{\ \rule{0.5em}{0.5em}}

\begin{document}

\title{Bound states due to a strong $\delta$ interaction
supported by a curved surface}
\author{P.~Exner$^{a,b}$ and S.~Kondej$^a$}
\date{}
\maketitle

\begin{quote}
{\small \em a) Nuclear Physics Institute, Academy of Sciences,
25068 \v Re\v z \\ \phantom{a) }near Prague, Czech Republic
\\
b) Doppler Institute, Czech Technical University,
B\v{r}ehov{\'a}~7, \\ \phantom{a) }11519 Prague, Czech Republic
\\
\phantom{a) }\texttt{exner@ujf.cas.cz}, \texttt{kondej@ujf.cas.cz}
}
\end{quote}

\begin{quote}
{\small {\bf Abstract.} We study the Schr\"odinger operator
$-\Delta -\alpha \delta (x-\Gamma )$ in $L^2(\R^3)$ with a
$\delta$ interaction supported by an infinite non-planar surface
$\Gamma$ which is smooth, admits a global normal parameterization
with a uniformly elliptic metric. We show that if $\Gamma $
is asymptotically planar in a suitable sense and $\alpha>0$ is
sufficiently large this operator has a non-empty discrete
spectrum and derive an asymptotic expansion of the eigenvalues in
iterms of a ``two-dimensional'' comparison operator determined by
the geometry of the surface $\Gamma$. }
\end{quote}


\section{Introduction}
\label{introd}
\setcounter{equation}{0}

The fact that a quantum particle localized in a curved infinitely
extended region can have bound states is known for more than a
decade -- cf.~\cite{ES, DE} and references therein. It was first
demonstrated for curved hard-wall strips and tubes. The analogous
problem in curved layers is more complicated and the existence of
curvature-induced bound states has been demonstrated only recently
\cite{DEK}. In addition, the sufficient conditions known so far
apply to particular classes of layers and lack therefore the
universal character of the ``one-dimensional'' existence result
noticed first in \cite{GJ}.

Another recent observation concerns the fact that the effect can
persist if the transverse Dirichlet condition confinement is
replaced by a weaker one. This is important if we want to apply
the conclusions to models of quantum wires and similar structures
in which the confinement is realized by a finite potential step at
an interface of two different semiconductor materials. As a
consequence, an electron can pass between two parts of a quantum
wire also by tunneling through the classically forbidden region
separating them.

To make the task more feasible, one can study the idealized
situation in which the structure is infinitely thin and the
Hamiltonian is formally written as $-\Delta -\alpha \delta
(x-\Gamma ),$ where $\alpha $ is a real parameter. It can be
interpreted as a limiting case of a transverse confinement by
a deep and narrow potential well, at least as long as the
codimension of the manifold $\Gamma$ is one. The existence
of a nontrivial discrete spectrum has been proven in this
setting if $\Gamma$ is a curve in $\mathbb{R}^2\,$ \cite{EI}
and $\mathbb{R}^3\,$ \cite{EK} which is asymptotically straight
but not a straight line and satisfies suitable regularity
conditions; the analogous result holds also for curved arrays of
point interactions \cite{Ex}. The argument is in all the listed
cases based on an explicit expression of the resolvent: one can
check that the curvature gives rise to  perturbation of the
straight-line Birman-Schwinger operator which is Hilbert-Schmidt
and of a definite sign.

As in the case of a hard-wall confinement the problem becomes more
complicated if the region to which the particle is localized is
generated by a surface in $\mathbb{R}^3$. The above mentioned
method does not generalize directly to such a situation, because
surfaces lack -- in distinction to curves -- a natural
parameterization which would allow us to ``iron'' them into a
plane. This is why we address the question in this paper using a
different method which will make it possible to establish the
existence of curvature-induced bound states provided the
attractive $\delta$ interaction in the Hamiltonian is strong
enough. The idea is borrowed from \cite{EY1, EY2} and is based on
a combination of Dirichlet-Neumann bracketing with the use of
suitable curvilinear coordinates in the vicinity of the surface
$\Gamma$, which is supposed to be asymptotically planar with a
uniformly elliptic metric. In this way we not only prove in
Theorem~\ref{assmut} the existence of a nontrivial discrete
spectrum, but also we obtain an asymptotic expansion of the
eigenvalues as $\alpha\to\infty$ in terms of a suitable
``two-dimensional'' comparison operator determined by the geometry
of $\Gamma$ -- cf.~(\ref{comp}) below. In the appendix we give
precise meaning to the above statement about the relation between
our Hamiltonian and the operator with a deep and narrow potential
well centered at the surface.


\section{Main results} \label{mresul}
\setcounter{equation}{0}

Let us start by summarizing our main results in more precise
terms. As we have said the Hamiltonians we will study are the
Schr\"{o}dinger operators with the a singular perturbations
supported by an infinite surface $\Gamma $ in $\R^{3}$
corresponding to the formal expression
\begin{equation} \label{D+dfor}
-\Delta -\alpha \delta (x-\Gamma )\,,
\end{equation}
where $\alpha>0 $ is independent of $x$. A general way to give
meaning to (\ref{D+dfor}) as a well-defined self-adjoint operator
(denoted by $H_{\alpha ,\Gamma }$) is to employ the sum of
quadratic forms; we will do that in Sec.~\ref{suqufo} below.

The form sum definition works under rather weak assumptions about
the regularity of $\Gamma$. For further purposes we have to
restrict the class of surfaces: the main results of the paper will
be derived for $\Gamma $ which is assumed to be
\begin{itemize}
\item  $C^{4}$ smooth and admitting a global normal
parameterization with a uniformly elliptic metric tensor -- see
(\ref{norpar}) and (\ref{estgmn_1}) below,
\item
without ``near-intersections'' -- assumption (a$\Gamma $1) of
Sec.~\ref{geomsu},
\item  asymptotically planar -- assumption (a$\Gamma $2) or a
stronger hypotesis (a$\Gamma 2^{\prime }$) of Sec.~\ref{bounst}.
\end{itemize}
Our goal is to investigate spectral properties of $H_{\alpha
,\Gamma }$ in the asymptotic regime when $\alpha $ is large. To
this aim we employ the comparison operator
\begin{equation} \label{comp}
S=-\Delta _{\Gamma }-\frac{1}{4}(k_{1}-k_{2})^{2}\,,
\end{equation}
where $\Delta _{\Gamma }$ is the Laplace-Beltrami operator on
$\Gamma $ and $k_{1},k_{2}$ are the principal curvatures of
$\Gamma$. If both $k_{1},k_{2}$ are identically zero, i.e. if
$\Gamma $ is a plane, it is easy to show that operator $H_{\alpha,
\Gamma }$ has purely absolutely continuous spectrum given by
$\sigma _{ac}(H_{\alpha ,\Gamma })=[- \frac{1}{4}\alpha
^{2},\infty )\: ;$ it is sufficient to employ separation of
variables and to use the spectrum of one-dimensional $\delta $
interaction. The aim of this paper is to prove that a ``local''
deformation of $\Gamma $ leads to existence of bound states for
large enough $\alpha .$ More precisely, we will show in
Theorem~\ref{assmut} that if $\Gamma $ is not a plane and
satisfies the above assumptions, then
\begin{itemize}
\item  the bottom of the essential spectrum does not lie below
$\epsilon(\alpha ),$ where $\epsilon(\cdot )$ is a function such
that $\epsilon(\alpha )\rightarrow -\frac{1}{4}\alpha ^{2}$ as
$\alpha \rightarrow \infty\,,$
\item  there exists at least one isolated point of the spectrum below the
threshold of the essential spectrum for all sufficiently large
$\alpha $, and moreover, the eigenvalues of $H_{\alpha ,\Gamma }$
have the following asymptotical expansion,
$$
\lambda _{j}(\alpha )=-\frac{1}{4}\alpha ^{2}+\mu
_{j}+\mathcal{O}(\alpha ^{-1}\log \alpha ) $$
as $\alpha\to\infty$, where $\mu _{j}$ is the $j$-th eigenvalue of
$S.$ The existence of a nonempty discrete spectrum alone can be
proven even without the uniform ellipticity assumption -- see
Remark~\ref{trialf}.
\end{itemize}

\noindent Let us remark that the existence of a global normal
parametrization is not necessary for derivation of the above
asymptotic formula -- cf.~\cite{Ex2}. It plays an important role,
however, when we prove that the discrete spectrum is non-empty by
comparison with a suitable two-dimensional Schr\"odinger operator.
An analogy with curved Dirichlet layers \cite{CEK} suggests that
bound states may exist even for some classes of surfaces $\Gamma$
which are not simply connected, but the proof of this conjecture
remains an open problem.


\section{Preliminaries} \label{prelim}
\setcounter{equation}{0}

\subsection{Geometry of the surface and its neighbourhood} \label{geomsu}

Let $\Gamma \subset \R^{3}$ be an infinite $C^{4}$ smooth surface
which admits a global normal parameterization (we refer to \cite{K}
for the geometric notions used below). This requires the existence
of a point $o\in \Gamma $ such that the exponential map $\exp
_{o}:T_{o}\Gamma \rightarrow \Gamma $ is a diffeomorphism. Given
orthonormal basis $\{e_{1}(o),e_{2}(o)\}$ in $T_{o}\Gamma $ we
introduce the map $\gamma \equiv \gamma _{o}:T_{o}\Gamma \cong
\R^{2} \rightarrow \Gamma $ defined by
\begin{equation} \label{norpar}
s\equiv (s_{1},s_{2})\rightarrow \exp
_{o}\left(\sum_{i}s_{i}e_{i}(o)\right),
\end{equation}
which determinates the said normal parameterization.

We denote by $g_{\mu \nu }$ the surface metric tensor in normal
coordinates, i.e. $g_{\mu \nu }=\gamma _{,\mu }\cdot\gamma _{,\nu
}$ and use the standard convention $g^{\mu \nu }=(g_{\mu \nu
})^{-1}.$ By means of the determinant $g:=\det g_{\mu\nu}$ we
define the invariant element of surface $d \Gamma =g^{1/2}d^{2}s$.
Furthermore, the tangent vectors $\gamma _{,\mu}$ are linearly
independent and their cross product $\frac{\gamma _{,\mu }\times
\gamma _{,\nu }}{\left| \gamma _{,\mu }\times \gamma _{,\nu
}\right| }$ defines the unit normal field $n(s)$ on $\Gamma .$

The extrinsic properties of surface can be characterized in terms
of the Weingarten tensor obtained by raising the index in the
second fundamental form,
\begin{equation} \label{Weigar}
h_{\mu }\ ^{\nu }:=-n_{,\mu }\cdot\gamma _{,\sigma }g^{\sigma \nu }\,.
\end{equation}
The eigenvalues of $h_{\mu }\ ^{\nu}$ are the pricipal curvatures
$k_{1},$ $k_{2}$; by means of them we define the Gauss curvature $K$
and mean curvature $M:$
\begin{equation} \label{defiKM}
K=\det h_{\mu }\ ^{\nu }=k_{1}k_{2}\,,\quad
M=\frac{1}{2}\mathrm{Tr\:} h_{\mu }\ ^{\nu
}=\frac{1}{2}(k_{1}+k_{2})\,.
\end{equation}
It follows immediately from the above formulas that
\begin{equation} \label{relaMK}
K-M^{2}=-\frac{1}{4}(k_{1}-k_{2})^{2}.
\end{equation}
Global quantities characterizing $\Gamma $ are the total Gauss
curvature
$
\mathcal{K}:=\int_{\Gamma }Kd\Gamma
$
which exists provided $K\in L^{1}(\Gamma ,d\Gamma )$ (if this is
not true the integral can sometimes exist as the principal value
with respect to the geodesic radius -- see Remark~\ref{polar}b
below) and the total mean curvature $\mathcal{M}:=
\left(\int_{\Gamma }M^{2}d\Gamma \right)^{1/2} $ (we set
$\mathcal{M}=\infty $ if $M\notin L^{2}(\Gamma ,d\Gamma )$).

The approximation methods which we will use in further discussion
force us to impose additional assumptions on $\Gamma$ which will
allow us to work in a neighbourhood of the surface. Given $\delta >0$
we consider the layer $\Omega _{\delta }$ built over $\Gamma $ and
defined by virtue of the map
\begin{equation} \label{mapLsu}
\mathcal{L}:\D_{\delta }\ni q\equiv (s,u)\rightarrow
\gamma (s)+un(s),\quad \D_{\delta }:=\{(s,u):s\in \R^{2},u\in
(-\delta ,\delta )\}.
\end{equation}
The fact that $\gamma $ is a diffeomorphism excludes automatically
the possibility of self-intersections of $\Gamma .$ Henceforth we
assume also that $\Gamma $ does not admit ``near-intersections'';
this is guaranteed by the following requirement:
 \begin{description}
 \item (a$\Gamma $1) there exists $d>0$ such that the map
 $\mathcal{L}:\D_{d}\rightarrow \Omega _{d}$ is injective.
 \vspace{-0.8ex}
 \end{description}
Let us stress that this is a restriction imposed on the {\em
global} geometry of $\Gamma$ which does not follow, e.g., from the
mere decay of the curvatures expressed by the assumption (a$\Gamma
$2) of Sec.~\ref{bounst} (although it is implied by (a$\Gamma
2^{\prime }$)). An example is easily constructed using
deformations of the plane in the form of a smooth `bubble' with a
narrow `bottleneck'; it is sufficient to consider a suitable array
of such deformations with properly changing parameters.

Using the parameterization (\ref{mapLsu}) we can find the metric
tensor of $\Omega _{d}$ regarded as a submanifold in $\R^{3}$,
$$
G_{ij}=
\left( \begin{array}{cc} (G_{\mu \nu }) & 0 \\
0 & 1 \end{array} \right)
,\quad G_{\mu \nu }=(\delta _{\mu }^{\sigma }-uh_{\mu }\ ^{\sigma })
(\delta _{\sigma
}^{\rho }-uh_{\sigma }\ ^{\rho })g_{\rho \nu }.
$$
In particular, the volume element of $\Omega _{d}$ is given by
$d\Omega := G^{1/2}d^{2}s\,du$, where $ G:=\det G_{ij}$ takes the
following form,
\begin{equation} \label{detGij}
G=g\left[ (1-uk_{1})(1-uk_{2})\right] ^{2}=g(1-2Mu+Ku^{2})^{2}.
\end{equation}
For the sake of brevity we employ the shorthand $\xi (s,u)\equiv
1-2M(s)u+K(s)u^{2}.$ Moreover, we will use the Greek notation for
the range $ (1,2)$ of indices  and the Latin for $(1,2,3).$ The
index numbering $(1,2,3)$ here refers to the coordinates
$(s_{1},s_{2},u).$

With a later purpose on mind we state some useful estimates.
Suppose that $k_{1},k_{2}$ are uniformly bounded (in fact we will
assume more -- see (a$\Gamma 2$) below) and put
$$ \varrho:=
(\{\max\left\| k_{1}\right\| _{\infty },\left\|
k_{2}\right\| _{\infty
}\})^{-1}. $$
Is easy to verify that for $d<\varrho $ the following inequalities
are satisfied in the layer neighbourhood $\Omega _{d},$
\begin{equation} \label{estixi}
C_{-}(d)\leq \xi \leq C_{+}(d),
\end{equation}
where $C_{\pm}(d):=(1\pm d\varrho^{-1})^{2}.$ Consequently, we
have
\begin{equation} \label{estGmn}
C_{-}(d)g_{\mu\nu}\leq G_{\mu\nu} \leq C_{+}(d)g_{\mu\nu}.
\end{equation}
To make use of the last inequality we need an information about
the metric of the surface. To prove our main result we will
require the uniform ellipticity of the metric tensor $g_{\mu
\nu}$, i.e. we suppose that there exist positive constants
$c^{+}\,,c^{-}$ such that
\begin{equation}  \label{estgmn_1}
c^{-}\delta_{\mu \nu }\leq g_{\mu \nu }\leq c^{+}\delta_{\mu \nu
}\,
\end{equation}
is satisfied as a matrix inequality.

\begin{remarks} \label{polar}
\rm{ (a) Combining (\ref{estixi}) and (\ref{detGij}) one can check
that the uniform boundedness of $k_{1},k_{2}$ (together with the
injectivity given by (a$\Gamma $1)) ensures that the map
$\mathcal{L}:\D_{d}\rightarrow \Omega _{d}$ is diffeomorphic if
$d<\varrho .$
\\ [.5em]
 (b)
By means of the change of variables $\phi $ given by
$$
s_{1}(r,\upsilon)=r\cos \upsilon ,\quad s_{2}(r,\upsilon)=r\sin \upsilon
$$
we pass to the geodesic polar coordinates (g.p.c.) $\,(r,\upsilon
)=(y_{1},y_{2}).$ In this notation $r=r(s)=(s_{1}^{2}
+s_{2}^{2})^{1/2}$ determines the geodesic radius. The metric
tensor
$$
\tilde{g}_{\mu \nu
}(y)=\sum_{\sigma \rho }\frac{\partial s_{\sigma }}{\partial y_{\mu }}\frac{%
\partial s_{\rho }}{\partial y_{\nu }}g_{\sigma \rho }(\phi (y))
$$
acquires in the g.p.c. the diagonal form  $\tilde{g}_{\mu
\nu}=\mathrm{diag} (1,\rho ^{2})$, where $\rho $ satisfies the
Jacobi equation
\begin{equation} \label{Jacobi}
\ddot{\rho}(r,\upsilon )+K(r,\upsilon )\rho (r,\upsilon )=0;\qquad\rho
(0,\upsilon )=0,\quad \dot{\rho}(0,\upsilon )=1.
\end{equation}
\\ [.5em]
(c)
There exist various sufficient conditions for (\ref{estgmn_1}).
For instance if $\Gamma $ is radially symmetric surface and $K\in L^{1}
(\Gamma ,d\Gamma )$ but $\mathcal{K} \neq 2\pi $ then it is easy to
show from (\ref{Jacobi}) that there exist positive constants $\tilde{c}^{+},
\tilde{c}^{-}$ such that
\begin{equation}  \label{estgmn}
\tilde {c}^{-}\tilde{g}_{\mu \nu }^{0}\leq \tilde{g}_{\mu \nu
}\leq \tilde{c}^{+}\tilde{g}_{\mu \nu }^{0}\,,
\end{equation}
where $\tilde{g}_{\mu \nu }^{0}$ is the metric tensor of a plane
in polar coordinates, in other words, $\tilde{g}_{\mu \nu
}^{0}=\mathrm{diag}(1,r^{2})$, which in turn implies
(\ref{estgmn_1}). This class of surfaces includes, for instance,
any hyperboloid of revolution. For surfaces without radial
symmetry a different sufficient condition for (\ref{estgmn_1}) is
needed, e.g., $\|K \|_{L^{1}(\Gamma ,d\Gamma)}< 2\pi$. For another
sufficient condition see Remark~\ref{rea1a2}b below.}
\end{remarks}


\subsection{Schr\"{o}dinger operators with singular perturbation
supported by the surface $\Gamma$ } \label{subsur}

\subsubsection{Construction of the Hamiltonian} \label{suqufo}

The Hamiltonians we will be interested in are Schr\"{o}dinger\
operators with perturbations supported by $\Gamma .$ A general way
to construct such operators is to employ the form sum technique.
Let us define the measure $\mu $ by
$$
\mu \equiv \mu _{\Gamma }:\mu _{\Gamma }(B):=\mathrm{vol}(B\cap \Gamma )
$$
for each Borel set $B$ in $\R^{3},$ where $\mathrm{vol}(\cdot )$
is two dimensional Hausdorff measure on $\Gamma .$ Using the fact
that the map $\gamma $ is a diffeomorphism and making use of
Theorem~4.1 in \cite{BEKS} is it easy to check that $\mu $ belongs
to the generalized Kato class. Consequently, the immbedding
operator
$$
I_{\mu}\psi =\psi,\quad I_{\mu}:\mathcal{S}(\R^{3})\subset W^{2,1}
(\R^{3})\rightarrow L^{2}(\R^{3},\mu)\equiv L^{2}(\mu )
$$
is continuous and it can be extended to whole space
$W^{2,1}(\R^{3}).$ Clearly, we have the natural identification
$L^{2}(\mu )\cong L^{2} (\R^{2},d\Gamma).$

With this prerequisite we are able to construct the mentioned
class of operators. Given $\alpha>0$ we define the quadratic form
$$
\eta _{\alpha }\left[ \psi \right] \equiv \eta _{\alpha }
(\psi ,\psi )=
(\nabla \psi,\nabla \psi)-\alpha \int_{\R^{3}
}\left|I_{\mu} \psi (x)\right| ^{2}d\mu (x),\quad \psi \in
 W^{2,1}(\R^{3})
$$
where $(\cdot,\cdot)$ stands for the scalar product in
$L^{2}(\R^{3}).$ It follows from Theorem~4.2 of \cite{BEKS} that
$\eta _{\alpha }$ is below bounded and closed; therefore the
associated operator $H_{\alpha , \Gamma }\equiv H_{\alpha }$ is
below bounded and self-adjoint in $L^{2}(\R^{3})$, and can be
interpreted as the self-adjoint realization of the symbol
(\ref{D+dfor}).

There is an alternative definition of $H_{\alpha }$ in terms of
boundary conditions on $\Gamma$; it is more illustrative because
it shows that in the direction transverse to the surface the
interaction is nothing else than a $\delta$ potential. For the
surface $ \Gamma $ with the properties specified in the previous
section we consider the Laplace operator
$$
(\dot{H}_{\alpha }\psi )(x)=-\Delta \psi (x)\,,\quad x\in
\R^{3}\setminus \Gamma \,,
$$
with the domain consisting of functions from $C_{0}(\R^{3})\cap
W^{2,2}(\R^{3}\backslash \Gamma )$ and having a jump of the normal
derivative on $\Gamma $ given by
$$
\left.\frac{\partial \psi }{\partial n}(x)\right|_+
-\left.\frac{\partial \psi }{\partial n }(x)\right|_- =-\alpha
\psi (x)\,.
$$
It is easy to verify that $\dot{H}_{\alpha }$ is e.s.a. and that
by the Green formula it reproduces the form $\eta _{\alpha }$;
therefore the closure of $\dot{H}_{\alpha }$ coincides with
$H_{\alpha }.$

\subsubsection{Approximation by scaled potentials }

Before proceeding further, let us say a few words about the
interpretation of the operator $H_{\alpha }.$ If $\Gamma$ is
smooth we can employ the standard approximation of the $\delta$
interaction by a family of squeezed potentials. To show this let
us consider again the layer neighbourhood $\Omega _{d}$ of $\Gamma
$ where $d<\varrho .$ Given $ W\in L^{\infty }(-1,1)$ we define
the scaled potentials with the support on $\Omega _{d}$, i.e.
$$ V_{d}(x):= \left\{
 \begin{array}{ccc} 0 & \quad \mathrm{if} \quad x & \notin \Omega _{d}  \\
-\frac{1}{d}W(\frac{u}{d}) & \quad \mathrm{if} \quad x &  \in
\Omega _{d} \end{array} \right.
$$
and associate with them the operators
$$
H_{d}(W):=-\Delta +V_{d}:D(\Delta)\rightarrow L^{2}(\R^{3}),
$$
where $-\Delta :D(\Delta)\rightarrow L^{2}(\R^{3})$ is the Laplace
operator. Since the potentials are bounded the operators
$H_{d}(W)$ are also self-adjoint with the domain $D(\Delta)$. This
family approaches $H_{\alpha }$ as $d\rightarrow 0$ in the
following sense:
\begin{theorem}  \label{approx}
Let the surface $\Gamma $ satisfies $(a\Gamma1)$. Then $H_{d}(W)
\rightarrow H_{\alpha }$ as $d \rightarrow 0,$ where $\alpha =\int_{-1}
^{1}W(t)dt$, in the norm-resolvent sense.
\end{theorem}
The proof is postponed to the appendix.

\subsubsection{Schr\"{o}dinger operators with the perturbation
on $\Gamma $ in the vicinity of the surface }

Let us return to our main subject. Our strategy is to estimate
(the negative spectrum of) the operator $H_{\alpha }$ using
operators acting in the layer neighbourhood of $\Gamma$. For the
set $\Omega _{d} $ we define the quadratic forms $\eta _{\alpha
}^{+}[\psi ],$ $\eta _{\alpha }^{-} [\psi ]$ with the domains
$D(\eta _{\alpha }^{+})=W_{0}^{2,1} (\Omega _{d}),$ $D(\eta
_{\alpha }^{-},)=W^{2,1}(\Omega _{d})$ which act as
$$ \left\| \nabla \psi (x)\right\| _{L^{2}(\Omega
_{d})}^{2}-\alpha \int_{\R^{3} }\left| I_{\mu }\psi (x)\right|
^{2}d\mu \,;$$
since the forms $\eta _{\alpha }^{\pm}$ are closed the associated
operators $H_{\alpha }^{\pm}$ are self-adjoint in $L^{2}(\Omega
_{d}).$ Now the Dirichlet-Neumann bracketing \cite{RS}
trick yields the estimate
\begin{equation} \label{bracke}
-\Delta _{\Sigma _{d}}^{N}\oplus H_{\alpha }^{-}\leq H_{\alpha
}\leq - \Delta _{\Sigma _{d}}^{D}\oplus H_{\alpha }^{+}\,, \quad
\Sigma_{d}\equiv \R^{3} \setminus \overline{\Omega }_{d}
\end{equation}
and as long we are interested in the negative point of spectrum, we
may take into account $H_{\alpha }^{\pm}$ only, because the
``exterior'' operators $\Delta _{\Sigma _{d}}^{D}$, $\Delta
_{\Sigma _{d}}^{N}$ are positive by definition.

It is useful to treat the estimating operators $H_{\alpha }^{\pm}$
in the coordinates $q= (s,u).$ We pass to the curvilinear
coordinates by means of the unitary transformation
$$
U\psi =\psi \circ \mathcal{L}:L^{2}(\Omega _{d})\rightarrow L^{2}
(\D_{d} ,d\Omega ).
$$
We denote by $(\cdot ,\cdot )_{G}$ the scalar product in the space
$ L^{2}(\D_{d},d\Omega ),$ then the operators $UH_{\alpha
}^{+}U^{-1},$ $UH_{\alpha}^{-}U^{-1}$ living in
$L^{2}(\D_{d},d\Omega )$ are associated with the forms $\psi
\mapsto \eta _{\alpha }^{+}[U\psi ],$ $\eta _{\alpha } ^{-}[U\psi
]$ having the value
\begin{equation} \label{foret1}
(\partial _{i}\psi ,G^{ij}\partial _{j}\psi )_{G}-\alpha
\int_{\Gamma }\left| \psi (s,0)\right| ^{2}d\Gamma\,,
\end{equation}
which differ by their domains, $W_{0}^{2,1}(\D_{d},d\Omega )$ and
$W^{2,1}(\D_{d}, d\Omega )$ for the $\pm$ sign, respectively.
Since the functions belonging to these spaces are not necessary
continuous the expression $\psi (s,0)$ in (\ref{foret1}) can be
given meaning using the trace mapping from
$W_{0}^{2,1}(\D_{d},d\Omega )$ or $W^{2,1}(\D_{d},d\Omega )$ to
$L^{2}(\Gamma ,d\Gamma ).$ For convenience we will use in the
following the same notations for $H_{\alpha }^{\pm},$ $\eta _
{\alpha }^{\pm},$ and its unitary ``shifts'' to the space
$L^{2}(\D_{d},d\Omega ).$

It is also useful to remove the factor $\xi $ from the weight
$G^{1/2}$ in space $L^{2}(\D_{d},d\Omega)$. This can be done
by means of another unitary transformation,
\begin{equation} \label{operaB}
\hat{U}\psi =\xi ^{1/2} \psi:L^{2}(\D_{d},d\Omega)\rightarrow
L^{2} (\D_{d},d\Gamma du)\,.
\end{equation}
We will denote the scalar product in $L^{2} (\D_{d},d\Gamma du)$
by $(\cdot,\cdot)_{g}$.

The operators
$$
B^{+}_{\alpha}:=\hat{U}
H^{+}_{\alpha}\hat{U}^{-1}\,, \quad B^{-}_{\alpha}:=\hat{U}H^{-}_{\alpha}
\hat{U}^{-1}
$$
acting in $ L^{2}(\D_{d},d\Gamma du)$ are associated with the
forms $b^{\pm}_{\alpha}$ given by $b^{+}_{\alpha}[\psi]:=\eta
^{+}_{\alpha}[\hat{U}^{-1}\psi]$ and $b^{-}_{\alpha}[\psi]:=\eta
^{-}_{\alpha}[\hat{U}^{-1}\psi]$. By a straightforward computation
we get
\begin{eqnarray*} \label{}
b^{+}_{\alpha}[\psi] &\!=\!&
(\partial_{\mu}\psi,G^{\mu\nu}\partial_{\nu}\psi)_{g}
+((V_{1}+V_{2})\psi,\psi)_{g}+\|\partial_{3}\psi \|_{g}^{2}-\alpha
\int _{\Gamma } |\psi (s,0)|^{2}d\Gamma\,, \\
b^{-}_{\alpha}[\psi]&\!=\!& b^{+}_{\alpha}[\psi]+\int _{\Gamma
}\zeta (s,d) |\psi(s,d)|^{2}d\Gamma -\int _{\Gamma }\zeta
(s,-d)|\psi(s,-d)|^{2} d\Gamma
\end{eqnarray*}
for $\psi$ from $W^{2,1}_{0}(\D_{d},d\Gamma du)$ and $W^{2,1}(\D_{d},
d\Gamma du)$, respectively, where $\zeta :=\frac{M-Ku}{\xi}$ and
\begin{equation} \label{poteV1}
V_{1}:=g^{-1/2}(g^{1/2}G^{\mu \nu}J_{,\nu})_{,\mu }+
J _{,\mu}G^{\mu \nu}J_{,\nu},\quad
V_{2}:=\frac{K-M^{2}}{\xi^{2}}, \quad J:=\frac{1}{2}\ln \xi .
\end{equation}


\section{Curvature-induced bound states}
\label{bounst}
\setcounter{equation}{0}

Let us turn now to the spectral analysis of $H_{\alpha }.$ The
main tool we will use to establish the existence of isolated
points of spectrum of $H_{\alpha }$ for large $\alpha $ is the
Dirichlet-Neumann bracketing (\ref{bracke}). The idea is to
construct operators $H_{\alpha }^{\pm}$ in the neighbourhood
depending on parameter $\alpha ,$ i.e. $d=d(\alpha ) $ such that
$d(\alpha ) \rightarrow 0$ as $\alpha \rightarrow \infty $, which
would provide us with a sufficiently exact spectral approximation
for $H_{\alpha }.$ As usual working with a minimax-type argument
we have to localize first the bottom of the essential spectrum.

\subsection{Essential spectrum}

Let us start with the case when $\Gamma $ is a plane in $\R^{3}.$
Then the translational invariance allows us to separate the
variables showing thus that
$$ \sigma (H_{\alpha })=\sigma _{ac}(H_{\alpha })=\left\lbrack
-\frac{ 1}{4}\alpha ^{2},\infty \right). $$
Next we assume that the surface $\Gamma $ admits a deformation
which is localized in the following sense:
 \begin{description}
 \item{(a$\Gamma $2)} $\;K,M \rightarrow 0$ as the geodesic radius
 $r\rightarrow \infty .$
 \vspace{-0.8ex}
 \end{description}
The main result of this part says that the bottom of
$\sigma_\mathrm{ess}(H_{\alpha })$ can be pushed down by the
deformation at most by a quantity which vanishes as $\alpha\to\infty$.
\begin{theorem}  \label{essspe}
Let $\alpha >0$ and suppose that the surface $\Gamma $ satisfies
$(a\Gamma 1)$, $(a\Gamma 2)$. Then
\begin{equation} \label{incess}
\sigma _{\mathrm{ess}}(H_{\alpha })\subseteq [\epsilon (\alpha
),\infty ),
\end{equation}
where $\epsilon (\cdot)$ is a function such that
$$
\epsilon (\alpha ) \rightarrow
-\frac{\alpha ^{2}}{4}\quad as \quad \alpha \rightarrow \infty .
$$
\end{theorem}
To prove the above theorem we will need a statement which follows
directly from Lemma~\ref{lemmEY} given below: there exist constants
$C,$ $C_{N}$ such that
\begin{equation} \label{bount-}
\int_{-d}^{d}\left| \frac{df(u)}{du}\right| ^{2}du-\alpha \left|
f(0)\right| ^{2}\geq \left(-\frac{\alpha ^{2}}{4}-C_{N}\alpha ^{2}
\mathrm{e}^{-\alpha d/2} \right)\|f\|^{2}_{L^{2}(-d,d)}
\end{equation}
with $f\in W^{2,1}(-d,d),$ holds for $\alpha >Cd^{-1}.$ \\ [0.5em]
\begin{proof}
Let us first note that the inclusion (\ref{incess}) is equivalent
to
\begin{equation} \label{infess}
\inf \sigma _{\mathrm{ess} }(H_{\alpha}) \geq \epsilon(\alpha ).
\end{equation}
In view of (\ref{bracke}) and positivity of  $-\triangle _{\Sigma
_ {d }}^{N}$ the inequality (\ref{infess}) follows from
\begin{equation} \label{estsi1}
\inf \sigma _{\mathrm{ess} }({H}_{\alpha }^{-})\geq \epsilon
(\alpha ),
\end{equation}
where $H_{\alpha }^{-}$ acts in $L^{2}(\Omega _{d}),\:
d<\varrho.$ Now we can proceed in analogy with the proof of
Theorem~4.1 in \cite{DEK}. Let us divide the surface $\Gamma $
into two components $\Gamma _{\tau }^{\mathrm{int}}:=\{s \in
\Gamma :r(s)<\tau \}$ and $\Gamma _{\tau}^{\mathrm{ext}}:=\Gamma
\backslash \overline{\Gamma }_{\tau }^{\mathrm{int}}.$ This gives
rise to the division of the layer neighbourhood to $\D _{\tau
}^{\mathrm{int}}$ and $\D _{\tau }^{\mathrm{ext}}$, where $\D
_{\tau }^{\mathrm{int}}:=\{(s,u)\in \D _{d}\,:s\in \Gamma _{\tau}
^{\mathrm{int} }\}$ and $\D _{\tau }^{\mathrm{ext}}:=\D
_{d}\backslash \overline{\D} _{\tau }^{\mathrm{int}}.$ Let us
consider the Neumann decoupled operators
$$ H^{-,\mathrm{int}}_{\alpha , \tau }\oplus
H^{-,\mathrm{ext}}_{\alpha , \tau}, $$
where $H^{-,\omega }_{ \alpha ,\tau},$ $\,\omega
=\mathrm{int},\mathrm{ext}$, are the operators associated with the
forms $\eta ^{-,\omega }_{ \alpha ,\tau}$ acting as (\ref{foret1})
and with the domains $W^{2,1}(\D ^{\omega }_ {\tau },d\Omega).$ Since
$H^{-}_{\alpha ,\tau}\geq H^{-,\mathrm{int}}_{\alpha , \tau
}\oplus H^{-,\mathrm{ext}} _{\alpha , \tau },$ and the spectrum of
$H^{-,\mathrm{int} }_{ \alpha ,\tau }$ is purely discrete \cite{D}
we obtain by the minimax principle
\begin{equation} \label{estsi2}
\inf \sigma _{\mathrm{ess} }({H}_{\alpha ,\tau }^{-})\geq \inf
\sigma _{\mathrm{ess} }(H^{-,\mathrm{ext} }_{ \alpha ,\tau }).
\end{equation}
Thus to verify the claim it suffices to check $\inf \sigma
_{\mathrm{ess} }({H}_{\alpha ,\tau }^{-,\mathrm{ext}})\geq
\epsilon(\alpha ).$ By the assumption (a$\Gamma $2) the quantities
$m_{\tau }^{+}:= \sup _{\Gamma ^{\mathrm{ext}}_{\tau } }\xi$ and
$m_{\tau }^{-}:= \inf _{\Gamma ^{\mathrm{ext}}_{\tau } }\xi $ tend
to one as $\tau \rightarrow \infty .$ Using (\ref{foret1}),
(\ref{bount-}), and the block form of $G^{ij}$ we get the estimate
\begin{eqnarray*}
\lefteqn{\eta ^{-,\mathrm{ext}}_{\alpha, \tau}[\psi ]\geq \int
_{\D^{\mathrm{ext}}_{\tau}}|
\partial _{3} \psi (q)|^{2}d \Omega -\alpha \int_{\Gamma _{\tau}^{\mathrm{ext}}}|
\psi (s,0)|^{2}d\Gamma }
\nonumber\\
 &&{} \geq m_{\tau }^{-}  \int _{\D^{\mathrm{ext}}_{\tau}}|
\partial _{3} \psi (q)|^{2}d\Gamma du-\alpha \int_{\Gamma _{\tau}^{\mathrm{ext}}}
|\psi (s,0)|^{2}d\Gamma \nonumber\\
 &&{} \geq \varepsilon _{\tau}\int
_{\D^{\mathrm{ext}}_{\tau}}| \psi (q)|^{2}d \Omega \nonumber\\
\end{eqnarray*}
where $ \varepsilon _{\tau}:=\frac{\alpha ^{2}}{m_{\tau} ^{+}
m_{\tau} ^{-}} [- \frac {1}{4}-C_{N}\exp (-\frac{1}{2}\frac{\alpha
}{m^{-}_{\tau}} d) ].$ Since $\tau $ is an arbitrary parameter we
get the inclusion (\ref{incess}) with $\epsilon (\alpha )=- \frac
{\alpha ^{2}}{4}-C_{N}\alpha ^{2}\mathrm{e}^{-\alpha d/2}.$
\end{proof}

\begin{remarks} \label{rea1a2}
\rm{ (a) The assumption (a$\Gamma 2$) can be replaced by a hypothesis
about the normal vector to $\Gamma$, namely
 \begin{description}
 \item(a$\Gamma 2^{\prime }$) $\;n\rightarrow n_0\;$ as
 $\,r\rightarrow \infty\,,$ where $n_0$ is a fixed vector.
 \vspace{-0.8ex}
 \end{description}
It is easy to see that the latter implies (a$\Gamma 2$), to this
end one has to combine (\ref {Weigar}) and (\ref{defiKM}). Of
course, the converse statement is not true: for example the
elliptic paraboloid satisfy (a$\Gamma 2$) but not (a$\Gamma
2^{\prime }$). As we will see in the further discussion the
assumption (a$\Gamma 2^{\prime }$) implies at the same time
(a$\Gamma 1$). Let us show that the claim of Theorem~\ref{essspe}
can be strengthened in this situation. Specifically, if $\alpha>0$
and the surface $\Gamma $ satisfies $(a\Gamma 2^{\prime })$, then
$$
\sigma _{\mathrm{ess}}(H_{\alpha })\subseteq \left\lbrack
-\frac{\alpha ^{2}}{4},\infty \right)\,. $$
To prove this we have to show that
\begin{equation} \label{infesn}
\inf \sigma _{\mathrm{ess}}(H_{\alpha })\geq -\frac{\alpha
^{2}}{4}\,.
\end{equation}
Similarly as in the proof of Theorem~\ref{essspe} we divide the
surface $\Gamma $ into two components $\Gamma _{\tau
}^{\mathrm{int}}$ and $\Gamma _{\tau }^{\mathrm{ext}}.$

Let us assume for a moment that $d(\tau )\equiv d_{\tau}:\R_{+}
\rightarrow \R_{+}$ is an arbirary function and consider one
parameter family of maps $\mathcal{L} _{\tau}:\D_{d_{\tau }}
\rightarrow \Omega _{d_{\tau }}.$ Using (a$\Gamma 2^{\prime }$)
and the mean value theorem we may show that the number ensuring
the injectivity of $\mathcal{L} _{\tau}$ in $\D^{\mathrm{ext}}_
{d_{\tau }}=\{(s,u):s\in \Gamma _{\tau}^{\mathrm{ext}},u\in
(-d_{\tau },d_{\tau })\}$ is proportional to the expression
$Q(\tau)\equiv \inf_{s_{1},s_{2}\in \Gamma _{\tau }^{\mathrm{ext}}}
\frac{|s_{1}-s_{2}| }{| n(s_{1})- n(s_{2})| } $ and
$$Q(\tau)
\rightarrow \infty \quad \mathrm{as} \quad   \tau \rightarrow
\infty \,. $$
This particularly means that the assumption (a$\Gamma 2^{\prime
}$) -- in distinction to the weaker hypothesis (a$\Gamma 2$) --
implies (a$\Gamma 1$) as we have indicated above. Furthermore,
under (a$\Gamma 2^{\prime }$) we can find $d_{\tau }$ such that
$d_{\tau }\rightarrow \infty $ as $\tau\rightarrow \infty $ and
the maps $\mathcal{L} _{\tau}$ are injective in
$\D^{\mathrm{ext}}_{d_{\tau }}.$ Relying on the same arguments as
in the previous proof we can show that to get (\ref{infesn}) it
suffices to check
$$ \inf \sigma _{\mathrm{ess}}(H_{\alpha ,d_{\tau
}}^{-,\mathrm{ext}})\geq -\frac{\alpha ^{2}}{4}\,, $$
where $H_{\alpha ,d_{\tau }}^{-,\mathrm{ext}}$ is associated with
the form $\eta _{\alpha ,d_{\tau }}^{-,\mathrm{ext}}$ acting as
(\ref{foret1}) and with the domain $W^{2,1}(\D_{d_{\tau }},d\Omega ).$
According to the previous discussion the assumption (a$\Gamma
2^{\prime }$)  implies $\sup_{\Gamma _{\tau
}^{\mathrm{ext}}}\left| M\right| ,\: \sup_{\Gamma _{\tau }
^{\mathrm{ext}}}\left| K\right| \rightarrow 0$, so we can always
choose $d_{\tau }$ such that $n_{\tau }^{+}\equiv \sup_{\D_{\tau
}^{\mathrm{ext}}}\xi $ and $n_{\tau }^{-}\equiv \inf_{\D_{\tau
}^{\mathrm{ext}}} \xi \rightarrow 1$ as $\tau \rightarrow \infty$
(recall that we can choose $d_\tau$ to go to infinity in an
arbitrarily slow way). Mimicking now the calculations from the
proof of Theorem~\ref{essspe} we arrive at
$$
\eta _{\alpha ,d_{\tau }}^{-,\mathrm{ext}}[\psi ]\geq
\tilde{\varepsilon}_ {\tau }\int_{\D_{d_{\tau }}^{\mathrm{ext}}}\left|
\psi (q)\right| ^{2}d\Omega \,, $$
where $\tilde{\varepsilon}_{\tau }:=\frac{\alpha ^{2}}{n_{\tau
}^{+}n_{\tau }^{-}}(-\frac{1}{4}-C_{N}\exp (-\frac{1}{2}\frac{
\alpha }{n_{\tau }^{-}}d_{\tau })).$ Since $\tilde{\varepsilon
}_{\tau } \rightarrow -\frac{\alpha ^{2}}{4}$ as $\tau \rightarrow
\infty $ and $\tau $ is an arbitrary parameter we get the stated
inequality (\ref{infesn}).
\\ [.5em]
(b) Notice that the assumption (a$\Gamma 2^{\prime }$) implies the
uniform ellipticity (\ref{estgmn_1}). Indeed, it means that to
each $\varepsilon>0$ there is a compact $\Sigma_\varepsilon$ such
that the inequality $|n(s)\!-\!n(s_0)|< \varepsilon$ holds for all
$s,s_0\in \R^2 \setminus \Sigma_\varepsilon$. Without loss of
generality we can fix $s_0$ and suppose that $n(s_0)= (0,0,1)$,
then $\Gamma$ coincides outside $\Sigma_\varepsilon$ with the
graph of a smooth function $f$. In that case we have explicit
expressions $n=g^{-1/2}(-f_1,-f_2,1)$ and
 $$ (g_{\mu\nu}) = \left(\begin{array}{cc} 1+f_1^2 & f_1f_2 \\
 f_1f_2 & 1+f_2^2 \end{array}\right)\,, $$
where $f_j\equiv \partial_j f$, which imply $\max\{ f_1,f_2\} \le
\varepsilon (1\!-\!\varepsilon^2)^{-1/2}$, and thus
(\ref{estgmn_1}) outside $\Sigma_\varepsilon$. On the other hand,
the eigenvalues of $(g_{\mu\nu})$ are continuous functions of the
parameters and thus they reach their maxima and minima in
$\Sigma_\varepsilon$. }
\end{remarks}


\subsection{Existence of bound states and asymptotics of
the eigenvalues } \label{bsasym}

In order to show the existence of bound states and to derive the
asymptotic behaviour of eigenvalues of $ H_{\alpha }$ we employ
the ``comparison" operator
$$
S:=-\Delta _{\Gamma }+K-M^{2}:D(-\Delta _{\Gamma })\rightarrow L^{2}
(\R^{2},d\Gamma ),
$$
where $-\Delta _{\Gamma }$ is the Beltrami-Laplace operator given
by
$$ -\Delta _{\Gamma }=-g^{-1/2}\partial_ {\mu }g^{1/2}g^{\mu
\nu }\partial _{\nu } $$
and $D(-\Delta _{\Gamma })$ is its
maximal domain in $L^{2}(\R^{2}, d\Gamma ).$ By (\ref{estgmn_1}) and
the assumption (a$\Gamma$2) it is straightforward to check that
$S$ is a well defined self-adjoint in $L^{2}(\R^{2},d\Gamma )$ and
its domain coincides with $W^{2,2} (\R^{2}).$ Moreover, using
(\ref{relaMK}) we infer that $K-M^{2}$ is an atractive potential
vanishing at infinity. Applying again (\ref{estgmn_1}) and the
standard results about two dimensional Schr\"{o}dinger operator
(see for example \cite{S}) we may conclude that $\sigma
_{\mathrm{ess}}(S)=[0,\infty )$, and that $S$ has at least one
negative eigenvalue provided $\Gamma$ is not a plane.

We denote by $\mu _{j}$ the $j$th eigenvalue of $S$ counted with
multiplicity, $j=1,\dots,N$ with $1\leq N \leq \infty $. Our main
result looks as follows.
\begin{theorem} \label{assmut}
Let $\Gamma $ satisfy assumptions $(a\Gamma 1)$, (\ref{estgmn_1})
and $(a\Gamma 2)$, or alternatively $(a\Gamma 2').$ Unless $\Gamma$
is a plane, there exists at least one isolated eigenvalue of
$H_\alpha$ below the threshold of the essential spectrum for
$\alpha$ large enough. Moreover, the eigenvalues $\lambda _{j}
(\alpha)$ of $H_{\alpha }$ have the following asymptotic expansion,
$$ \lambda _{j}(\alpha)=-\frac{1}{4} \alpha ^{2}+\mu
_{j}+\mathcal{O}(\alpha ^{-1}\log \alpha ) \quad \mathrm{as} \quad
\alpha\to\infty\,. $$
\end{theorem}

\bigskip

\noindent As the first step to prove the theorem we derive some
useful estimates. Consider again the layer neighbourhood $\Omega
_{d}$ of $\Gamma$ and quadratic forms $\eta _{\alpha}^{\pm}$
acting as (\ref{foret1}) with the domains $W_{0}^{2,1} (\D
_{d},d\Omega)$ and $W^{2,1}(\D_{d},d\Omega)$, respectively.
As we have already mentioned, due to the Dirichlet-Neumann bracketing the
associated operators $H_{\alpha }^{+}$, $H_{\alpha }^{-}$ living
in $L^{2}(\D_{d},d\Omega )$ give the upper and lower estimates for
negative eigenvalues of $H_{\alpha }.$ Moreover, in view of the
unitary equivalence of $H_{\alpha }^{\pm}$ and  $B_{\alpha
}^{\pm}$ we can consider $B_{\alpha }^{\pm}$ instead of $H_{\alpha
}^{\pm}.$

Since $B^{\pm}_{\alpha}$ are still not quite easy to handle, our
next aim to estimate them by operators with separated variables.
Using the explicit form of the potential $V_{1}$, and taking into
account (\ref{estixi}), (\ref{estGmn}), (\ref{estgmn_1}) together
with assumption (a$\Gamma $2) we can find the numbers $v^{+}$,
$v^{-}$ such that
\begin{equation} \label{estiV1}
dv^{-}\leq V_{1}\leq dv^{+}
\end{equation}
holds for $d< \varrho .$ On the other hand, we can estimate the
potential $V_{2}$ by
\begin{equation} \label{estiV2}
C_{-}^{-2}(d)(K-M^{2})\leq V_{2} \leq C_{+}^{-2}(d)(K-M^{2})\,,
\end{equation}
where $C_\pm$ were introduced in (\ref{estixi}). Now we define
another pair of estimating operators in $ L^{2}(\R^{2},d\Gamma
)\otimes L^{2}(-d,d)$ by
\begin{equation} \label{Btilda}
\tilde{B}^{+}_{\alpha ,d}:=U^{+}_{d } \otimes 1 + 1\otimes T^{+}
_{\alpha ,d}\,, \quad \tilde{B}^{-}_{\alpha ,d}:=U^{-}_{d }\otimes
1 + 1\otimes T^{-}_{\alpha ,d}\,,
\end{equation}
where
$$
U_{d}^{\pm }=-C_{\pm}(d)\Delta_{\Gamma }+C_{\pm}^{-2}(d)(K-M^{2})
+v\pm d
$$
and $T^{\pm}_{\alpha ,d}$ are associated with the quadratic forms
given by
\begin{eqnarray*} 
 t^{+}_{\alpha ,d}[\psi] &\!=\!& \int_{-d}^{d}|\partial_{3}\psi
|^{2}du - \alpha |\psi (0)|^{2}\,, \\ t^{-}_{\alpha ,d}[\psi]
&\!=\!& \int_{-d}^{d}|\partial_{3}\psi |^{2}du - \alpha |\psi
(0)|^{2}-D_{d}(|\psi(d)|^{2}+|\psi(-d)|^{2})
\end{eqnarray*}
for $\psi\in W^{2,1}_{0}(-d,d)$ and $W^{2,1}(-d,d)$, respectively.
The coefficient $D_{d}:=2(\left\| M\right\| _{\infty }+\left\|
K\right\| _{\infty }d)$ coming from the Neumann boundary
conditions is in distinction to the similar quantity for
$B^-_\alpha$ independent of the surface variables $s$. It is clear
from (\ref{estiV1}) and (\ref{estiV2}) that $\tilde{B}_{\alpha
,d}^{\pm }$ give the sought bounds for $B_{\alpha }^{\pm },$ i.e.
\begin{equation} \label{estimB}
B_{\alpha }^{+}\leq \tilde{B}_{\alpha ,d}^{+}\quad \mathrm{and}
\quad \tilde{B}_{\alpha ,d}^{-} \leq B_{\alpha }^{-}.
\end{equation}
Since $\tilde{B}_{\alpha ,d}^{\pm}$ have separated variables their
spectra express through those of their ``constituent'' operators.
Consider first the ``surface'' part. Repeating the arguments we
have used for the operator $S$ we may conclude that $\sigma
_{d}(U_{d}^{\pm })\neq \emptyset .$ Our aim is now to find the
asymptotic behaviour of the eigenvalues $\mu _{j}^{\pm }(d)$ of
$U_{d}^{\pm }$ for small $d$, or more precisely, we would like to
show that numbers $\mu _{j}(d)$ approach the eigenvalues $\mu
_{j}$ of the comparison operator $S$ as $d\rightarrow 0.$
\begin{lemma}  \label{appmuj}
The eigenvalues of $U^{\pm}_{d}$ have the following asymptotics
\begin{equation} \label{asymmu}
\mu _{j}^{\pm }(d)=\mu _{j}+ C_{j}^{\pm}d+\mathcal{O}(d^{2})
\end{equation}
for $d\rightarrow 0,$ where $C_{j}^{\pm}$ are constants.
\end{lemma} \label{asymud}
\begin{proof}
Assume $d <\varrho .$ Applying the explicit form for $U_{d}^{+}$
we immediately get
\begin{equation} \label{Ud-CdS}
U_{d}^{+}-C_{+}(d)S=v^{+}d+(C_{+}^{-2}(d)-C_{+}(d))
(K-M^{2}).
\end{equation}
Using the formula for $C_{+}(d)$ it is easy to check that any
function $d\mapsto m(d)$ with the asymptotic behaviour
$$
(v^{+}+(\left\| K\right\| _{\infty }+\left\| M\right\|
_{\infty }^{2})\varrho^{-1})d+\mathcal{O}(d^{2})
$$
as $d\rightarrow 0,$ gives the upper bound for the r.h.s. of (\ref{Ud-CdS})
for sufficiently small $d$.
Thus we get the following estimate
$$
\|U_{d}^{+}-C_{+}(d)S\|\leq m(d).
$$
Combining this inequality and the minimax principle we get
$$
|\mu_{j}^{+}(d)-C_{+}(d)\mu_{j}|\leq m(d),
$$
which implies
$$
|\mu_{j}^{+}(d)-\mu_{j}|\leq m(d)+d|(2\varrho ^{-1}+\varrho ^{-2}
d)\mu _{j}|.
$$
and thus (\ref{asymmu}) for $\mu ^{+}_{j}(d).$ The proof for $\mu
^{-}_ {j}(d)$ is analogous.
\end{proof}

\bigskip

Let us pass to the transverse part. To estimate the negative
eigenvalues of $T_{\alpha ,d}^{\pm }$ we will use the lemma
borrowed from \cite{EY1}.
\begin{lemma} \label{lemmEY}
There exist positive constants $C,\: C_{N}$ such that each of the
operators $T_{\alpha ,d}^{\pm }$ has a single negative eigenvalue
$\kappa _{\alpha ,d}^{\pm }$ satisfying
$$ -\frac{\alpha ^{2}}{4}-C_{N}\alpha ^{2}\mathrm{e}^{-\alpha d/2}
 <\kappa _{\alpha ,d}^{-} < -\frac{\alpha ^{2}}{4}<
 \kappa _{\alpha ,d}^{+}<-\frac{\alpha ^{2}}{4}%
+2\alpha ^{2}\mathrm{e}^{-\alpha d/2} $$
for any $\alpha >C\max \{d^{-1},D_{d}\}.$
\end{lemma}

Now we are ready to prove Theorem~\ref{assmut}. Put
\begin{equation} \label{dalpha}
d=d(\alpha ):=6\alpha ^{-1}\log \alpha .
\end{equation}
From Lemma~\ref{lemmEY} we know that operators $T_{a,d(\alpha
)}^{\pm }$ have single negative eigenvalues $\kappa _{\alpha
}^{\pm }\equiv \kappa _{\alpha ,d(\alpha )}^{\pm }$ and in view of
the decomposition (\ref{Btilda}) we infer that $\kappa
_{\alpha}^{\pm }+\mu _{j,\alpha }^{\pm },$ where $\mu _{j,\alpha
}^{\pm } =\mu _{j}^{\pm }(d(\alpha )),$ are eigenvalues of
$\tilde{B}_{\alpha }^{\pm }.$ Using again Lemma~\ref{lemmEY}
together with Lemma~\ref {appmuj} we may conclude that $\kappa
_{\alpha}^{\pm }+\mu _{j}^ {\pm }(\alpha )$ have the following
asymptotics
$$
\kappa _{\alpha}^{\pm }+\mu _{j,\alpha }^{\pm }=-\frac{1}{4} \alpha
^{2}+\mu _{j}+\mathcal{O}(\alpha ^{-1}\log \alpha )
$$
Finally, it follows from (\ref{bracke}) and (\ref{estimB}) that the same
asymptotics holds for eigenvalues of $H_{\alpha}.$
This completes the proof.

\begin{remark} \label{trialf}
\rm{The existence of isolated points of spectrum below the
threshold of the essential spectrum for large $\alpha $ can be
alternatively proven by constructing a trial function $\psi \in
D(H_{\alpha })$ such that the following inequality
$$ (H_{\alpha }\psi ,\psi)<\left(-\frac{\alpha ^{2}}{4}-D
\right)\|\psi\|^2, $$
holds with a positive $D$. By Dirichlet bracketing it suffices to
find a function $f\in D(H^{+}_{\alpha })$ satisfying
\begin{equation} \label{triaf1}
(H_{\alpha }^{+}f ,f)_{G}<\left(-\frac{\alpha ^{2}}{4}-D \right)
\|f\|^2_{G}.
\end{equation}
Using the unitary equivalence $H_{\alpha }^{+}$ and  $B_{\alpha }
^{+}$ and inequality $B_{\alpha }^{+}\leq \tilde{B}_
{\alpha ,d}^{+}$ we infer that (\ref{triaf1}) is satisfied if
\begin{equation} \label{triaf2}
(\tilde{B}_{\alpha ,d}^{+}h ,h)_{g}+\frac{\alpha
^{2}}{4}\|h\|^2_{g}<-D\|h\|^2_{g}\,, \quad h=\hat{U}f\,
,
\end{equation}
holds. Let $h$ be a radially symmetric function given by
$h(r,u)=\phi (r)\chi (u)$ where $\phi$ is an arbitrary for a
moment and $\chi $ is the normalized eigenfunction of
$T^{+}_{\alpha ,d}$ corresponding to the negative eigenvalues
$\kappa ^{+}_{\alpha ,d}$ (see Lemma \ref{lemmEY}). Substituting
again $d$ from (\ref{dalpha}) one can show that there exist
functions $\theta _{1}(\cdot),$ $\theta _{2}(\cdot)$ and $\beta (\cdot )$
such that $\theta _{1}(\alpha ),\theta _{2}(\alpha )\rightarrow 1,$
$\beta (\alpha )\rightarrow 0$ as $
\alpha \rightarrow \infty $ and the expression
\begin{equation} \label{pushdo}
\theta_{1}(\alpha )\left\| \nabla \phi \right\| ^{\prime 2}_{g}+
\beta (\alpha )\left\| \phi \right\| ^{\prime 2}_{g}+\theta _{2}
(\alpha )((K-M^{2})\phi ,\phi )^{\prime }
_{g}
\end{equation}
gives the upper bound for the l.h.s. of (\ref{triaf2}). Here we
use the notation $( \cdot ,\cdot )^{\prime }_{g}$ for the scalar
product in $L^{2}(\R^{2},d\Gamma ).$ Given $r_{0}>0$ we define
$$
\phi _{\sigma }(r):=\min \left\{ 1,\frac{K_{0}(\sigma r)}{K_{0}(\sigma r_{0})%
}\right\} \,,\quad \sigma >0\,,
$$
where $K_{0}$ is the Macdonald function, and substitute $\phi
_{\sigma }$ for $\phi $ in the above formulae. Since $K_{0}$ is
strictly decreasing the function $\phi _{\sigma }$ is not smooth
at $r=r_{0}$, and consequently, $\phi _{\sigma }\chi$ does not
belong to the domain of operator $\tilde{B}^{+}_ {\alpha ,d}.$
However, it belongs to $W^{2,1}_{0}(\D_{d},d\Gamma du)$ which
coincides with the domain of the form associated with $\tilde{B}
^{+}_{\alpha ,d}.$ Therefore it suffices to show that the quantity
(\ref{pushdo}) (with the replacement $\phi\mapsto\phi_{\sigma}$)
is smaller than $-D\left\| \phi _{\sigma }\right\| ^{\prime
2}_{g}.$ Using the properties of the Macdonald function one can
show that $ \left\| \nabla \phi _{\sigma }\right\| _{g}^ {\prime
}\rightarrow 0$ and $\phi_{\sigma }(r)\rightarrow 1-$ pointwise as
$\sigma \rightarrow 0$-- cf.~\cite{EV}. Moreover, since the
expression $K-M^{2}$ is strictly negative in some open subset of
$\R^{2}$ there exist positive constants $\sigma _{0},\: \alpha
_{0}$ and $C=C(\sigma _{0},\alpha _{0})$ such that
$$
\theta_{1}(\alpha )\left\| \nabla \phi _{\sigma _{0}}\right\|
^{\prime 2}_{g}+ \theta _{2}(\alpha )((K-M^{2})\phi _{\sigma _{0} },\phi
 _{\sigma _{0}})^{\prime }_{g}<-C
$$
holds for all $\alpha >\alpha _{0}.$ Choosing $\tilde {D}<C
\|\phi _{\sigma _{0}}\|^{\prime -2}_{g}$ and using the fact that
$\beta (\alpha )\rightarrow 0$ as $\alpha \rightarrow \infty $
we can always find $\tilde{\alpha }>\alpha _{0}$ such that
the inequality
%
$$
\theta _{1}({\alpha })\left\| \nabla \phi _{\sigma _{0}}\right\|
^{\prime 2}_{g}+\beta ({\alpha})\left\| \phi _{\sigma _{0}}\right\|
^{\prime 2}_{g}+\theta _{2} ({\alpha })((K-M^{2})\phi _{\sigma_{0}},
\phi _{\sigma _{0}})_{g}^{\prime }<-\tilde{D}\left\| \phi _{\sigma_{0}}
\right\| ^{\prime 2}_{g}
$$
holds for all $\alpha \geq \tilde{\alpha}.$ Putting $D=\tilde{D}$
we get the claim. Let us stress that the above argument does not
require the assumption (\ref{estgmn_1}) about the uniform
ellipticity of the metric tensor $g_{\mu\nu}$.}
\end{remark}


\setcounter{section}{1}
\setcounter{equation}{0}
\renewcommand{\theequation}{\Alph{section}.\arabic{equation}}
\renewcommand{\theclaim}{\Alph{section}.\arabic{equation}}
\section*{Appendix: Proof of Theorem~\ref{approx}}

Since the argument is analogous to that of Theorem 4.1 in
\cite{EI} we here present only the main steps. Given $z\in
\varrho(-\Delta )$ we denote by $\mathrm{R}^{0}(z)$ the free
resolvent, $\mathrm{R}^{0}(z)=(-\Delta -z)^{-1}$, which is an
integral operator with the kernel
\begin{equation}  \label{formGk}
\mathrm{G}_{z}(x,y)=\frac{1}{(2\pi )^{3}}\int_{\R^{3}}\frac{
e^{ip(x-y)}}{p^{3}-z}=\frac{1}{4\pi }\frac{e^{i\sqrt{z}\left| x-y\right| }}{
\left| x-y\right| }.
\end{equation}
It suffices for further discussion to consider only a subset of
$\varrho(-\Delta )$, for instance, the negative halfline. Thus we
put $z=k^{2},$ where $k=i\kappa$ with $\kappa >0$ and introduce
the notations $R^{0}(k)=\mathrm{R}^{0}(k^{2}),$
$G_{k}=\mathrm{G}_{k^{2}}.$

First we express the resolvent of $H_{d}(W)$ in the Birman-Schwinger form
\begin{eqnarray} \label{resoRd}
\lefteqn{ R^{d}(k) :=(H_{d}(W)-k^{2})^{-1} }  \\ && =
R^{0}(k)+R^{0}(k)V_{d}^{1/2}[I+\left| V_{d}\right|
^{1/2}R^{0}(k)V_{d}^{1/2}]^{-1}\left| V_{d}\right| ^{1/2}R^{0}(k),
\nonumber
\end{eqnarray}
for $k^{2}\in \varrho(-\Delta )\cap \varrho(H_{d}(W))$, with the
usual $V_{d}^{1/2}=\left| V_{d}\right| ^{1/2}\mathrm{sgn}\,
V_{d}.$ Mimicking the analysis in \cite{EI} the second term at the
r.h.s. of (\ref{resoRd}) can be written as the product of integral
operators $B_{d}(I-C_{d})^{-1} \tilde{B}_{d}$ mapping
$L^{2}(\R^{3})\rightarrow L^{2}(\D_{1},d\Gamma dt)\rightarrow
L^{2}(\D_{1},d\Gamma dt)\rightarrow L^{2}(\R^{3})$ with the
kernels
\begin{eqnarray*}
B_{d}(x;s^{\prime },t^{\prime }) &\!=\!& G_{k}(x-x^{\prime
}(s^{\prime },dt^{\prime }))\xi (s^{\prime },dt^{\prime
})W(t)^{1/2}\,, \\ \tilde{B}_{d}(s,t;x^{\prime }) &\!=\!& \left|
W(t)\right| ^{1/2}\xi (s,dt)G_{k}(x^{\prime }-x(s,dt))\,, \\
C_{d}(s,t;s^{\prime },t^{\prime }) &\!=\!& \left| W(t)\right|
^{1/2}G_{k}(x(s,dt)-x^{\prime }(s^{\prime },dt^{\prime
}))W(t^{\prime })^{1/2}\,,
\end{eqnarray*}
where $x(s,dt)\equiv \gamma (s)+dtn(s).$ Using (\ref{formGk}) we
estimate $ \left\| C_{d}\right\| \leq \left\| W\right\| _{\infty
}\left| k\right| ^{-2}; $ thus $\left\| C_{d}\right\| <1$ holds,
uniformly in $d$, for all sufficiently large $\kappa .$ Thus we
can expand the operator under consideration as a geometric series,
\begin{equation} \label{expaBd}
B_{d}(I-C_{d})^{-1}\tilde{B}_{d}=\sum\nolimits_{j=0}^{\infty }B_{d}C_{d}^{j}%
\tilde{B}_{d}.
\end{equation}
Let us turn to the analysis of the resolvent of $H_{\alpha }.$ To
this aim we need the embedding operators associated with
$R^{0}(k)$, namely
$$ R_{\mu }^{0}(k):=I_{\mu }R^{0}(k):L^{2}(\R^{3})\rightarrow
L^{2}(\mu ),
$$
with the adjoint $\left[ R_{\mu }^{0}(k)\right] ^{\ast }:L^{2}(\mu
)\rightarrow L^{2}(\R^{3})$, and
$$
R_{\mu \mu }^{0}(k):=I_{\mu }\left[ R_{\mu }^{0}(k)\right] ^{\ast
}:L^{2}(\mu )\rightarrow L^{2}(\mu ).
$$
Using this notation one can express the resolvent of $H_{\alpha }$
as
\begin{eqnarray}
\lefteqn{ R(k) :=(H_{a}-k^{2})^{-1}}  \label{resolR} \\ &&
=R^{0}(k)+\alpha \left[ R_{\mu }^{0}(k)\right] ^{\ast }[I-\alpha
R_{\mu \mu }^{0}(k)]^{-1}R_{\mu }^{0}(k) \nonumber
\end{eqnarray}
for any $k^{2}\in \varrho(-\Delta )\cap \varrho(H_{\alpha })$ --
cf.~\cite{BEKS}. Using $\alpha =\int_{-1}^{-1}W(t)\,dt$ we can
expand the second term at the r.h.s. of ( \ref{resolR}) as
\begin{equation} \label{expaHa}
\sum\nolimits_{j=0}^{\infty }BC^{j}\tilde{B}\,,
\end{equation}
where $BC^{j}\tilde{B}$ is the product of operators mapping $
L^{2}(\R^{3})\rightarrow L^{2}(D_{1},d\Gamma dt)\rightarrow
L^{2}(D_{1},d\Gamma dt)\rightarrow L^{2}(\R^{3})$ with the kernels
\begin{eqnarray*}
B(x;s^{\prime },t^{\prime }) &\!=\!& G_{k}(x-\gamma (s^{\prime
}))W(t)^{1/2}\,, \\ \tilde{B}(s,t;x^{\prime }) &\!=\!& \left|
W(t)\right| ^{1/2}G_{k}(x^{\prime }-\gamma (s))\,, \\
C(s,t;s^{\prime },t^{\prime }) &\!=\!& \left| W(t)\right|
^{1/2}G_{k}(\gamma (s)-\gamma (s^{\prime }))W(t^{\prime })^{1/2}.
\end{eqnarray*}
Applying (\ref{expaBd}) and (\ref{expaHa}) and repeating the
argument of \cite{EI} we can estimate
\begin{eqnarray}
\lefteqn{ \left\| R^{d}(k)-R(k)\right\|  \leq
\sum\nolimits_{j=0}^{\infty }\left\|
BC^{j}\tilde{B}-B_{d}C_{d}^{j}\tilde{B}_{d}\right\| } \nonumber \\
&& \phantom{A} \le c(\| B_{d}-B\| +\left\|
\tilde{B}_{d}-\tilde{B}\right\| +\| C_{d}-C\| )
\label{noRd-R}
\end{eqnarray}
for a positive $c$. The first norm at the r.h.s. of (\ref{noRd-R})
can be estimated by
$$ \left\| W\right\| _{\infty }^{1/2}\left\{ \left\|
\breve{R}_{d,1}(k)-\breve{R}_{0}(k)\right\| +d(2\left\| M\right\|
_{\infty }+d\left\| K\right\| _{\infty })\left\|
\breve{R}_{d,1}(k)\right\| \right\} , $$
where $\breve{R}_{d,1}(k),\breve{R}_{0}(k)$ are the integral
operators acting from $L^{2}(\D_{1},d\Gamma dt)$ to
$L^{2}(\R^{3})$ with the kernels $G_{k}(x-x^{\prime }(s^{\prime
},t^{\prime }))$ and $G_{k}(x-\gamma (s^{\prime })).$ Since
$\left\| \breve{R}_{d,1}(k)\right\| $ is uniformly bounded w.r.t.
$d$, the norm convergence in question for $ d\rightarrow 0$ will
follow from the corresponding property of $\left\|
\breve{R}_{d,1}(k)- \breve{R}_{0}(k)\right\| $. The latter can be
estimated by the Schur-Holmgren bound,
$$
\left\| \breve{R}_{d,1}(k)-\breve{R}_{0}(k)\right\| \leq (h_{1}h_{\infty
})^{1/2},
$$
where
\begin{eqnarray*}
h_{\infty }&\!:=\!& \sup_{x\in \R^{3}}\int_{\D_{1}}\left| (
\breve{R}_{d,1}(k)-\breve{R}_{0}(k))(x,x^{\prime }(s^{\prime
},dt^{\prime }))\right| d\Gamma ^{\prime }dt^{\prime }\,, \\
h_{1}&\!:=\!&\sup_{x^{\prime }\in \D_{1}}\int_{\D_{1}}\left| (
\breve{R}_{d,1}(k)-\breve{R}_{0}(k))(x,x^{\prime })\right| dx\,.
\end{eqnarray*}
Finally, using the mean value theorem in combination with the fact
that $\left| G_{k}^{\prime }\right| \in L^{1}(\R^{3},g^{1/2}dx)$, where
$\left| G_{k}^{\prime }(x)\right|=\frac{1}{4\pi }\frac{(\kappa
\left| x\right| +1)}{ x^{2}}\, \mathrm{e}^{-\kappa \left| x\right|
}$ one can establish existence of a
constant $c_{1}$ such that $h_{1},h_{\infty }$ entering the
SH-bound are both majorized by the expression
$c_{1}d\left\|G_{k}^{\prime }\right\| _{L^{2}(\R^{3},g^{1/2}dx)}$.
The convergence of $ \left\| \tilde{B}_{d}-\tilde{B}\right\| $ and
$\left\| C_{d}-C\right\|$ is checked in the same way.

\bigskip


\subsection*{Acknowledgment}

The authors thank D.~Krej\v{c}i\v{r}\'{\i}k and the first referee
for useful comments. The research has been partially supported by
GA AS under the contract A1048101.

 \end{document}